\documentclass[twocolumn,showpacs,aps,prx,amsmath,amssymb]{revtex4}
\usepackage{bm,color}
\usepackage{ulem}

\newcommand{\beq}{\begin{equation}}
\newcommand{\eeq}{\end{equation}}
\newcommand{\bqa}{\begin{eqnarray}}
\newcommand{\eqa}{\end{eqnarray}}

\newcommand{\erf}[1]{Eq.~(\ref{#1})}

\newcommand{\ket}[1]{ |{#1} \rangle}

\newcommand{\Ket}[1]{\left|{#1}\right\rangle}

\newcommand{\mg}{\color{magenta}}
\newcommand{\blk}{\color{black}}
\newcommand{\blu}{\color{blue}}
\definecolor{ngreen}{rgb}{0.2,0.6,0.2}
\newcommand{\grn}{\color{ngreen}}
\definecolor{golden}{rgb}{0.8,0.6,0.1}
\newcommand{\gold}{\color{golden}}

\renewcommand\grn{\blk}
\renewcommand\blu{\blk}

\renewcommand\mg{\blk}
\renewcommand\gold{\blk}

\begin{document}
\title{Does nonlinear metrology offer improved resolution? \\ Answers from quantum information theory}
\author{Michael J. W. Hall and Howard M. Wiseman}
\affiliation{Centre for Quantum Dynamics, Griffith University, Brisbane, QLD 4111, Australia}

\begin{abstract}
A number of authors have suggested that nonlinear interactions can enhance resolution of phase shifts beyond the usual Heisenberg scaling of $1/n$, where $n$ is a measure of resources such as the number of subsystems of the probe state or the mean photon number of the probe state.  These suggestions are based on calculations of `local precision' for particular nonlinear schemes.  However, we show  that there is no simple connection between the local precision and the average estimation error for these schemes, leading to a scaling \blu puzzle. \blk This is partially resolved by a careful analysis of  iterative  implementations of the suggested nonlinear schemes.  However, it is shown that 
\grn the suggested nonlinear schemes are still limited to an exponential scaling in $\sqrt{n}$. \gold (This scaling may be compared to 
the exponential scaling in $n$ which is achievable if multiple passes are allowed, even for linear schemes.)   \blk  The question of whether nonlinear schemes may have a scaling advantage in the presence of loss is left open.   Our  results are based on a  new  bound for average estimation error  that depends on (i) an entropic measure of the degree to which the probe state can encode a reference phase value, called the $G$-asymmetry, and (ii) any prior information about the phase shift. This bound  is asymptotically stronger than bounds based on  the  variance of the phase shift generator.    The $G$-asymmetry is also shown to  directly bound   the average information gained per estimate.    
Our results \blu hold for any prior distribution of the shift parameter, and \blk generalise to estimates of any shift generated by an operator with discrete eigenvalues.

\end{abstract}

\maketitle



\section{Introduction}

In many measurement scenarios, an environmental variable acts to translate or shift a property such as the optical phase or position of a probe state. Accurate estimation of the  shifted  parameter allows a correspondingly accurate measurement of the environmental variable.  For example, interferometric measurements of  quantities ranging from temperature to  gravitational wave amplitudes rely on the estimation of an optical phase shift.  An important aim of quantum metrology is to determine the fundamental bounds on the resolution of such estimates, and how these bounds scale with available resources such as energy  \cite{glm2004,helstrom, holevo, WisMil10}.

Let us denote the initial probe state by the density operator $\rho_0$, and the generator of shifts by some Hermitian operator $G$. Then if the shift parameter $\Phi$ has 
the value $\phi$, the final probe state is  $\rho_{\phi} = \exp(-iG\phi)\rho_0\exp(iG\phi)$. In the following,  particular attention will be paid to  the estimation of a  {\it phase}  shift parameter, as this is sufficient for discussing various nonlinear estimation schemes previously proposed in the literature \cite{luis1,luis2,roy,boixo1,ramsey,boixo2,boixo3}.  The generator $G$ in this case has integer eigenvalues, 
so that $\rho_{\phi+2\pi}=\rho_\phi$. \blu More generally, however, our results apply  to any shift generator $G$ having a discrete eigenvalue spectrum. \grn This includes the atomic \blu scheme proposed in \cite{nap1}, which has recently led to the first experimental \grn demonstration of nonlinear quantum \blu metrology \cite{nap2}. \blk

\grn Returning to an optical example, \blk a linear phase shift of a single-mode optical probe state corresponds to $G=N$ where $N$ is the photon number operator.  Similarly, for a probe state comprising $m$ such modes, each undergoing a nonlinear quadratic phase shift, one has $G=(N_1)^2+\dots +(N_m)^2$ \cite{luis1,luis2}.  In cases like this, we quantify the resources $n$ by the total mean photon number $\sum_j\left< N_j\right>$.   Alternatively, for a  probe  comprising  $n$ atomic qubits, each with a Pauli  Z operator $\sigma^{(j)}_z$,   one may  consider  the  generator $G= \sigma^{(1)}_z+\dots +\sigma^{(n)}_z$, and powers thereof, corresponding to linear and nonlinear Ramsey interferometers respectively \cite{boixo1,ramsey,boixo2,boixo3}.  Again, $n$ quantifies the resources. 

We note that this quantification of resources $n$ is different from the $N$  (which we will denote ${\cal N}$)  used in  Ref.~\cite{higgins,berry,Higgins09,rapcom,tsang}.  The $n$ used here typically corresponds to the  conspicuous 
 physical resources required to generate the probe state,  and is what has
previously been used to claim an advantage when using nonlinear interactions \blu \cite{luis1,luis2,roy,boixo1,ramsey,boixo2,boixo3,nap1,nap2}.  \blk

If $\hat{\Phi}$ denotes an estimate of  a  shift parameter $\Phi$, for some measurement scheme, then a standard measure \gold \cite{vantrees,refnote}
of the performance of the estimate is given by the average estimation error (called rms error in Ref.~\cite{vantrees}), \blu
\begin{equation} \label{error}
\epsilon(\hat{\Phi}):= \grn \sqrt{E[{(\hat{\Phi}-\Phi)^2}]} , 
\end{equation}
\grn where the expectation value here is defined as
\beq
 E[{(\hat{\Phi}-\Phi)^2}] \equiv  {\int d\phi\, d\hat{\phi}\,(\hat{\phi}-\phi)^2p(\hat{\phi}|\phi)\wp(\phi) }  .
\eeq
Here \blu $p(\hat{\phi}|\phi)$ is the probability density of the estimate conditioned on a fixed shift value $\Phi=\phi$, and $\wp(\phi)$ denotes the prior probablity density of the shift parameter. \blk  Measurement schemes which minimise the average estimation error, for given resources such as the average photon number or number of qubits available, are of fundamental interest \grn in quantum metrology. \blk   

However, attention has often focused instead on minimising a different quantity, the `local precision', defined for a {\it fixed} value of the shift parameter,  $\Phi=\phi$, by \cite{braun,annphys}
\begin{equation} \label{dis}
P_\phi(\hat{\Phi}) :=  \left\langle\, \left( \frac{ \hat{\Phi}}{| d\langle \hat{\Phi}\rangle_\phi/d\phi|} -\phi \right)^2\, \right\rangle^{1/2}_{\grn \phi} \, ,
\end{equation}
\blu where $\langle\bullet\rangle_{\grn \phi}$ denotes an average with respect to the conditional probability density $p(\hat{\phi}|\phi)$. \blk Some proposed nonlinear measurement schemes can achieve local precisions which scale in terms of the number of resources $n$ as, for example,  $n^{-3/2}$ \blu \cite{luis1,boixo1,ramsey, nap1,nap2}, \blk $n^{-2}$ \cite{boixo2},  or $2^{-n}$ \cite{roy}, for some value of $\phi$.  Even so, it will be shown below that  the corresponding average estimation errors can scale no better than  the usual Heisenberg scaling, $n^{-1}$. 


 For estimates which are,  approximately, locally  unbiased for {\it all} values of $\Phi$ over some interval \cite{unbiased}, one has \blu 
\begin{equation} \label{unbiased}
\int d\phi\,\wp(\phi)\,P_\phi(\hat{\Phi}) \approx \epsilon(\hat{\Phi}) 
\end{equation}
\blk for shift parameters confined to this interval,  providing a simple connection to the average estimation error. However,  \grn many \blk phase estimates \grn are unbiased \blk only over very limited ranges \cite{future}, where these  ranges  are  of widths comparable to  the local precision itself.   Thus, for example, while a high local precision of $2^{-{ K}}$ in some region may allow the ${ K}$th binary digit of a phase shift to be estimated, it often will not allow the preceding digits to be estimated with any accuracy.  These must either already be known \mg (e.g., in phase tracking \cite{durk} or phase sensing \cite{xiang} applications),  \blk
\grn which requires the prior probability distribution $\wp(\phi)$ to be almost as narrow as the posterior distribution 
(after the measurement),  or they must be \blk determined using further resources.  Hence, \grn unless the phase is already very well known, \blk the scaling of $P_\phi(\hat{\Phi})$ may be a very poor guide to the scaling of $\epsilon(\hat{\Phi})$. 

 Indeed, whereas the local precision has a scaling  lower  bound set by the rms variance, $\Delta G$, of the generator for the probe state  \cite{braun,annphys}, the average estimation error   has an asymptotically stronger  (i.e. higher) lower  bound, set by the  entropy, $H(G)$, of the generator  \cite{hallwise,nair}. Thus, maximising  the variance, rather than the entropy, of $G$,  will not typically minimise $\epsilon(\hat{\Phi})$.   Here we further generalise and strengthen this entropic bound,  in Secs.~II and~III, to replace  $H(G)$ by the so-called $G$-asymmetry of the probe state \cite{vacc1}. The fundamental role of this quantity is emphasised by showing that it also bounds  the mutual information between the  shift parameter  $\Phi$ and any estimate $\hat{\Phi}$.   An important consequence demonstrated in Sec.~IV is that,  in  \blu a surprising \blk contrast to the case of local precision,   simply replacing $G$ by some nonlinear function thereof, such as $F=G^2$, cannot improve the average estimation error nor the information gain.

A careful analysis  in Sec.~V  shows that nonlinearity can improve the scaling of $\epsilon(\hat{\Phi})$ beyond $n^{-1}$ 
for  {\it iterative} implementations. 
 These are implementations where the  shift is applied on a sequence of probes of different sizes,   
so that $G$ is  replaced by a suitable sum of nonlinear generators.  
However, for  a probe state comprising $n$ qubits,  it is shown that even  adaptive variable-pass  implementations of previously proposed nonlinear schemes can at best achieve scalings exponential in $\sqrt{n}$ for the average estimation error. In contrast,  in Sec.~VI,  we show  that the best possible scaling for $\epsilon(\hat{\Phi})$ is exponential in $n$, both for qubit and optical probes, whether or not the generator is linear or nonlinear.  Moreover, an exponential scaling is in fact achievable via {\it linear} estimation schemes,  if 
{\it multipass} implementations are allowed.   
  Whether  nonlinear schemes are more robust than linear schemes to the presence of loss  is  left as a question for future investigation.


\section{An information bound}

   The mutual information between the  shift parameter and its estimate, $H(\hat{\Phi}:\Phi)$, is a measure of performance in its own right, quantifying the average number of bits obtained per estimate.  A general upper bound for mutual information is obtained here, applicable to {\it any} generator $G$ having a discrete spectrum, which will be used in Sec.~III below to obtain a lower bound for the average estimation error.  Several useful properties of this bound are also established.

 Consider a parameter $\Phi$ with some prior distribution ${ \wp(\phi)}$, and define 
an average prior state $\overline{\rho} = \int d\phi\,{ \wp(\phi)}\, \rho_\phi$. 
Then, using the Holevo bound \cite{nielsen}, one immediately has
$H(\hat{\Phi}:\Phi) \leq 
 S(\overline{\rho}) - \int  d\phi\, { \wp(\phi)}\, S(\rho_\phi)=S(\overline{\rho}) - S(\rho_0)$.  Here
$S(\rho) = - {\rm tr}[\rho\ln\rho]$ denotes the von Neuman entropy of the  state  $\rho$.   Now define 
\begin{equation} \label{ug} 
{\cal U}_{ G}(\rho) := \sum_g \Pi_g\rho\, \Pi_g = \lim_{w\rightarrow \infty} \frac{1}{w}\int_0^w d\phi\,e^{-iG\phi}\rho \, e^{iG\phi}, 
\end{equation} 
where $\Pi_g$ is the projection on to the eigenspace corresponding to eigenvalue $g$ of $G$, and the second equality may be checked by considering a basis diagonal in $G$. This map is unital, i.e., it maps the unit operator to itself. Using the nondecreasing property of von Neumann entropy under unital maps \cite{nielsen}, together with ${\cal U}_G(\overline{\rho})={\cal U}_G(\rho_0)$, then yields the desired upper bound,
\begin{equation}
\label{inf}
H(\hat{\Phi}:\Phi)\leq  A_G(\rho_0) := S({\cal U}_G(\rho_0))-S(\rho_0) ,
\end{equation} 
for the mutual information.

The upper bound, $A_G(\rho_0)$ in Eq.~(\ref{inf}), may be recognised as the increase in quantum entropy due to an ideal measurement of $G$ on the probe state, with post-measurement state ${\cal U}_G({\rho_0})$.  This entropy increase is relevant to bounding efficiencies in quantum thermodynamics \cite{lloyd}.   More generally, $A_G(\rho)$ represents the {\em asymmetry} of the state $
\rho$ with respect to a unitary group $G$ (in this paper, the  one-parameter  Abelian group with  Hermitian 
generator $G$) \cite{vacc1}. The $G$-asymmetry quantifies the degree to which $\rho_0$ can break the symmetry of $G$ (in this paper, the extent to which it carries information about  
the variable $\Phi$ which is conjugate to $G$) \cite{vacc1,vacc2,spekk,vacc3}. For the case where $G$ has integer eigenvalues, 
$A_G(\rho_0)$ quantifies to what extent $\rho_0$ can act as a phase reference, an attribute clearly essential for detecting phase shifts. 
Note that for $G$ with incommensurate eigenvalue gaps, the corresponding group is non-compact, but the above expression 
(\ref{ug}) for ${\cal U}_{ G}$ allows one to  generalise  the $G$-asymmetry (\ref{inf})  to  this case also.


A form of Eq.~(\ref{inf}) has been previously given for the special case  of compact groups where the average prior state $\bar\rho$ is symmetric with respect to the group \cite{vacc2,spekk,vacc3}. For the case of a phase-shift (i.e. a $G$ with integer eigenvalues) this means a prior distribution ${ \wp(\phi)}$ which is uniform over the unit circle.  Equation (\ref{inf}) represents a generalisation,  for the case of one-parameter groups,  to  an arbitrary  discrete  generator $G$  and arbitrary prior distributions of the shift parameter.   Note that $\Phi$  ranges over $(-\infty,\infty)$ if $e^{-iG\phi}$ is nonperiodic,  corresponding to a noncompact group. 

Several useful properties of $A_G(\rho)$ will be required further below.  First, if $\rho$ is pure and/or $G$ is nondegenerate, then the states $\rho_g=\Pi_g\rho\Pi_g/p_g$, with $p_g={\rm tr}[\rho\Pi_g]$, are pure and mutually orthogonal, and Eqs.~(\ref{ug}) and (\ref{inf}) yield
\begin{equation} \label{pure}
A_G(\rho) = H(G|\rho) - S(\rho), 
\end{equation}
where $H(G|\rho)= - \sum_g p_g \ln p_g$ is the entropy of the generator for state $\rho$. 
Second, one has the general bounds
\begin{equation} \label{bounds}
A_{f(G)}(\rho) \leq A_G(\rho) \leq H(G|\rho),
\end{equation}
\begin{equation} \label{convex}
A_G(\lambda\rho+(1-\lambda)\rho') \leq \lambda A_G(\rho) +(1-\lambda)A_G(\rho')
\end{equation}
where $f(G)$ is any function of $G$ and $0\leq\lambda\leq 1$.  The lower bound in Eq.~(\ref{bounds}) is saturated when $f$ is 1:1, as may be seen by replacing $G$ by $f(G)$ and $f$ by $f^{-1}$, while the upper bound is saturated for any pure state from Eq.~(\ref{pure}). The convexity of $A_G(\rho)$, as per Eq.~(\ref{convex}), implies that the $G$-asymmetry is maximised for pure states.

The lower bound in (\ref{bounds}) is obtained by noting that the eigenspaces of $G$ are subspaces of the eigenspaces of $f(G)$, so that ${\cal U}_G  \circ {\cal U}_{f(G)} = {\cal U}_G$, and using the nondecreasing property of von Neumann entropy under unital maps \cite{nielsen} for the particular case ${\cal U}_{f(G)}(\rho) \rightarrow  ({\cal U}_G  \circ {\cal U}_{f(G)})(\rho)$. To obtain the upper bound, let $|\psi\rangle$ be some purification of $\rho$ on a tensor product of the probe Hilbert space with an ancilla $a$, so that $\rho= {\rm tr}_a  [|\psi\rangle\langle\psi|]$.  Rewriting $A_G(\rho)$ as $\sum_g p_g S(\rho\|\rho_g)$, where $S(\rho\|\sigma)={\rm tr}[\sigma(\ln \sigma -\ln \rho)]$ denotes the relative entropy of  $\rho$ and $\sigma$, one then has
\begin{eqnarray*} 
H(G|\rho) &=& H(G\otimes 1|\,|\psi\rangle\langle\psi|) = A_{G\otimes 1}(|\psi\rangle\langle\psi|)\\ 
&=& \sum_g p_g S(|\psi\rangle\langle\psi|\,\|(\Pi_g\otimes 1)|\psi\rangle\langle\psi| (\Pi_g\otimes 1)/p_g)\\
&\geq& \sum_g p_g S(\rho\|\rho_g) = A_g(\rho)
\end{eqnarray*}
as desired, where the second equality follows from Eq.~(\ref{pure}), and the inequality from the nonincreasing property of relative entropy under the operation of tracing over the ancilla \cite{nielsen}.  Finally, Eq.~(\ref{convex}) may be obtained via the representation
$A_G(\rho) = \lim_{w\rightarrow\infty} w^{-1} \int_0^w d\phi\,S(\rho\| \rho_\phi)$,
following from Eq.~(\ref{ug}), and using the joint convexity property of the relative entropy \cite{nielsen}.

Equations~(\ref{inf}) and (\ref{bounds}) imply in particular that the mutual information is bounded by the entropy of the generator for the probe state, i.e., 
\begin{equation} \label{infbound}
H(\hat{\Phi}:\Phi)\leq H(G|\rho_0) .
\end{equation} 
Thus,  for example, for a generator having $d$ distinct eigenvalues, no more than $\ln d$ nats, \blu i.e., $\log_2 d$ bits, \blk of information  can be extracted per probe state about the value of the shift parameter.

\section{Bounds for resolution of shift parameters}
 
\subsection{Average estimation error}

 A strong bound for the average estimation error in Eq.~(\ref{error})  may be derived analogously to weaker bounds obtained in \cite{nair,yuen92}, i.e., by combining a quantum upper bound  --- such as Eq.~(\ref{inf}) ---  for the mutual information with the classical lower bound \cite{cover, rate}
\begin{equation} \label{lower}
H(\hat{\Phi}:\Phi)\geq H(\Phi)-\frac{1}{2}\ln [2\pi e\,\epsilon(\hat{\Phi})]\, ,
\end{equation} 
where  $H(\Phi) = - \int { \wp(\phi)} \ln[{ \wp(\phi)}] d\phi$ denotes the entropy of the prior probability density ${ \wp(\phi)}$ for $\Phi$. This lower bound is well known from rate-distortion theory, and follows from the inequality chain \cite{rate} 
$H(\hat{\Phi}:\Phi)=H(\Phi)-H(\Phi|\hat{\Phi})=H(\Phi)-H(\Phi-\hat{\Phi}|\hat{\Phi}) \geq H(\Phi)-H(\Phi-\hat{\Phi})
\geq H(\Phi)-\frac{1}{2}\ln [2\pi e\,\epsilon(\hat{\Phi})]$, where $H(A|B)$ denotes the conditional entropy $H(AB)-H(B)$.

In particular, the combination of Eqs.~(\ref{inf}) and (\ref{lower}) immediately yields the fundamental bound

\begin{equation} \label{ebound}
\epsilon(\hat{\Phi})  \geq (2\pi e)^{-1/2}\, e^{H(\Phi)}\,  \,e^{-A_G(\rho_0)} 
\end{equation}
 for the average estimation error, for any discrete generator $G$.   This bound both strengthens and generalises previous entropic bounds in the literature \cite{yuen92,rapcom,hallwise,nair}.   For example, Nair \cite{nair} and Yuen \cite{yuen92} use weaker upper bounds for the mutual information, corresponding to replacing $A_G(\rho_0)$ in Eq.~(\ref{ebound}) by the quantum channel capacity under a fixed photon number constraint.  Hall and co-workers have previously obtained  bounds in a different manner, based on entropic uncertainty relations, which correspond to replacing $A_G(\rho)$ in Eq.~(\ref{ebound}) by the upper bound in Eq.~(\ref{bounds})  \cite{rapcom,hallwise}  (and, alternatively, by the upper bound in Eq.~(\ref{pure}) for nondegenerate generators  \cite{hallwise}),  and replacing $e^{H(\Phi)}$ by $1/q_{\max}$, where $q_{\max}$ denotes the maximum value of ${ \wp(\phi)}$  \cite{hallwise}.  

Note that  our bound (\ref{ebound}) is  applicable to  iterative schemes, including adaptive ones,  where the  measurement performed  on some probe state components is dependent  (in practice, through additional known phase rotations)  on the outcomes of earlier measurements on other components \cite{WisMil10}.  This is because such a measurement scheme is formally equivalent to first applying shift generators $G_1, G_2,\dots$ to respective probe components (e.g., qubits or optical modes), corresponding to applying the total generator $G=G_1+G_2+\dots$, and then  
performing the measurements sequentially (and adaptively).  


\subsection{Local precision}

A  bound for the local precision in Eq.~(\ref{dis}) follows via the quantum Cramer-Rao inequality \cite{helstrom,holevo}, and has the form \cite{braun,annphys}
\begin{equation} \label{dbound}
P_\phi(\hat{\Phi}) \geq (2\Delta G)^{-1}\, ,
\end{equation}
where $\Delta G$ denotes the root mean square deviation of the  (total)  generator $G$ for the probe state.  
Note that, taking the averages of $\nu$ independent estimates, one obtains the usual statistical enhancement factor of $1/\sqrt{\nu}$ for both of the bounds in (\ref{ebound}) and (\ref{dbound}).  This will not be discussed further here, other than to remark that  although   the latter bound  for $P_\phi(\hat{\Phi})$ \gold may be \blk asymptotically achievable as $\nu\rightarrow \infty$ \gold \cite{braun,annphys,future,durk}, \blk  this does not imply anything about the achievability of the corresponding bound for $\epsilon(\hat{\Phi})$.  

\subsection{Comparisons}

 For generators with integer eigenvalues, i.e., phase shift generators,  the scaling  of $\epsilon(\hat{\Phi})$ with the exponentiated $G$-asymmetry  $ e^{-A_G(\rho_0)}$  in Eq.~(\ref{ebound}) implies a scaling 
with the root mean square error $\Delta G$ which is at least as strong as that for $P_\phi(\hat{\Phi})$ in Eq.~(\ref{dbound}) (ignoring multiplicative constants of order unity). This is  a consequence of the inequality chain
\begin{equation} \label{strict}
 e^{-A_G(\rho)} \geq e^{-H(G| \rho)}  \geq (2\pi e)^{-1/2}[(\Delta G)^2 + 1/12]^{-1/2} 
\end{equation}
for such generators,  where the first inequality follows from Eq.~(\ref{bounds}) and the second is well known \cite{rate,disc}.  
For the case of a completely unknown phase shift, with ${ \wp(\phi)}=1/(2\pi)$, Eqs.~(\ref{ebound}) and (\ref{strict}) yield the asymptotic lower bound 
$\epsilon(\hat{\Phi}) \gtrsim (e\Delta G)^{-1}$
for the average estimation error, which is comparable to the lower bound (\ref{dbound}) for local precision.  However, importantly, the bound in Eq.~(\ref{ebound}) is significantly  more powerful, as we now show. 

Consider, for example,  a probe comprising $n$ qubits, and generator
$G=\sigma^{(1)}_z+\dots +\sigma^{(n)}_z$.  The pure probe state $(\Ket{z,z,\dots,z} + \Ket{-z,-z,\dots,-z})/\sqrt{2}$, where $|\pm z\rangle$ denotes the eigenstates of $\sigma_z$, then gives the maximum possible value $\Delta G=n$ in Eqs.~(\ref{dbound}) and (\ref{strict}).  That is why this state, equivalent to a NOON state or  GHZ state \cite{noon}, is often considered in quantum metrology. However, the  corresponding $G$-asymmetry follows via Eq.~(\ref{pure}) as only $A_G(\rho_0)=H(G|\rho_0)=\ln 2$,  implying via Eq.~(\ref{ebound}) that the average estimation error does not decrease at all as a function of $n$.  The only way an average estimation error scaling as $n^{-1}$ would be possible from this state would be if there was sufficient prior information. That is, from Eq.~(\ref{ebound}), 
 if $-H(\Phi) + A_G(\rho_0)$ was of order $\ln n$. But since  $A_G(\rho_0)=\ln 2$, this means $-H(\Phi) \sim \ln n$ itself. Hence the amount of prior information about the parameter to be estimated would already be sufficient to locate it with the precision 
 achievable by the measurement.  

It is  thus apparent  that the lower bounds (\ref{ebound}) and (\ref{dbound}) can exhibit markedly different behaviour, with the former bound having an asymptotically stronger scaling in general.  It  follows that probe states generating optimal scaling for $P_\phi(\hat{\Phi})$, obtained by maximising $\Delta G$ under various constraints for some value of $\phi$, do not necessarily correspond to optimal bounds for $\epsilon(\hat{\Phi})$.  Since it is the latter quantity which has direct operational significance for the performance of the estimate, this has crucial implications for some nonlinear estimation schemes proposed in the literature, as will be seen below.


\blu
\section{Resolution puzzle: ~~~ nonlinearity vs $G$-asymmetry}
\blk
\subsection{Probes comprising $n$ qubits}

Several nonlinear phase estimation schemes have been proposed for Ramsey interferometry, based on a probe state comprising $n$ qubits \cite{boixo1,ramsey,roy,boixo2,boixo3}.  For example, defining $J_z:=\sigma^{(1)}_z+\dots + \sigma^{(n)}_z$,  the generator $(J_z)^{ q}$ has been considered in \cite{boixo1, ramsey, boixo3} for ${ q}=2,3,\dots$, and the generator $nJ_z$ in \cite{boixo2}.  A nonlinearity 
equivalent to $J_z^2$, although defined in terms of a fixed number of bosons shared between two modes, was considered in 
\cite{luis1}.  Further, the generators $H$ and $A$ defined via 
\begin{equation} \label{ha}
H+iA = \left(\sigma^{(1)}_x+ i\sigma^{(1)}_y\right)\otimes \dots \otimes \left(\sigma^{(n)}_x+ i\sigma^{(n)}_y\right)
\end{equation}
were considered in \cite{roy}.

Now, the linear generator $G=J_z$ has $n+1$ distinct eigenvalues, $-n, -n+2, \dots,n$, and hence  $A_G(\rho)\leq H(G|\rho)\leq \ln (n+1)$.  It follows immediately from Eq.~(\ref{ebound}) (see also Eq.~(22) of \cite{hallwise}), that the corresponding average estimation error can scale no better than $(n+1)^{-1}$ with qubit number, corresponding asymptotically to the Heisenberg scaling limit of $n^{-1}$.  

However, noting  from Eq.~(\ref{bounds}) that  $A_{f(G)}(\rho)\leq A_G(\rho)$,  for any function $f$, precisely the same conclusion follows for the nonlinear generators $G=(J_z)^{ q}$ and $G=nJ_z$.  That is, the average estimation error cannot achieve better than Heisenberg scaling for these generators.  Moreover, the nonlinear generators $G=H$ and $G=A$ of Eq.~(\ref{ha}) do not even allow the possibility of Heisenberg scaling, as  they each only have 3 distinct eigenvalues, 0 and $\pm 2^{n-1}$ \cite{royeig}.

The above results are in stark contrast to the best possible scalings of local precision for these schemes, which improve on Heisenberg scaling, with $n^{-{ q}}$ for $G=(J_z)^{ q}$  \cite{luis1,boixo1, ramsey, boixo3};   $n^{-2}$ for $G=nJ_z$ \cite{boixo2}; and $2^{-n}$ for $G=H$ or $A$ \cite{roy}.   This difference in scalings immediately raises a \blu conundrum: \blk how can nonlinearity improve the local precision, yet not the average estimation error? 

This \blu puzzle \blk may be further \blu deepened \blk by noting that the probe states yielding {\it optimal} local precisions are generally  an equally-weighted  superposition of two orthogonal eigenstates of $G$, corresponding to the maximum and minimum eigenvalues of the generator \cite{boixo1,ramsey,boixo2,boixo3}.  Thus, $A_G(\rho_0) = \ln 2$ for such a probe state,  implying that the average estimation error cannot decrease with $n$ at all, as discussed  in Sec.~III~C above.   Indeed,  Eq.~(\ref{infbound}) implies that no more than 1 bit of information about the phase shift can be gained via such an `optimal' probe state.

\subsection{Probes comprising optical modes}

An analogous puzzle holds for optical probes.  As a simple example, if $G=N$ is the number operator for a single mode field, then $H(G|\rho)\leq \ln (e\langle N+1\rangle)$, implying from Eqs.~(\ref{bounds}) and (\ref{ebound}) that the average estimation error can scale no better than $\langle N+1\rangle^{-1}$ (see also \cite{hallwise,yuen92,nair,rapcom}).  But for any nonlinear generator $G=f(N)$,  $A_G(\rho)\leq A_N(\rho)\leq H(N|\rho)$ from Eq.~(\ref{bounds}).  Hence, the same scaling bound also applies to nonlinear generators  for single mode fields.   

In contrast, nonlinearity can significantly enhance the local precision.  For example, choosing the coherent probe state $\rho_0=|\alpha\rangle\langle\alpha|$ and nonlinear generator $G=N^2$,  $P_\phi(\hat{\Phi})$  can scale as  $\langle N\rangle^{-3/2}$ for large $\langle N\rangle$  \cite{luis1}.  For this  case  the  photon number distribution is Poissonian, which is well approximated by a Gaussian distribution for large  $\langle N\rangle$. Thus,  using Eq.~(\ref{pure}),  $A_G(\rho_0)=H(N^2|\rho_0)=H(N|\rho_0)\approx (1/2)\ln (2\pi e \langle N\rangle)$,  implying via Eq.~(\ref{ebound}) that the corresponding average estimation error  can decrease with  $\langle N\rangle$.  However, it cannot scale even as well as the Heisenberg limit, $\langle N\rangle^{-1}$, but rather is lower bounded by the standard quantum limit scaling, $\langle N\rangle^{-1/2}$.  

These examples again \blu lead to the puzzle \blk that while nonlinearity can improve the scaling of the local precision, it cannot, by itself, influence the scaling of the average estimation error. This \blu raises \blk the question:  can  nonlinear schemes offer any   advantage over linear schemes?  

\blu
\section{Puzzle resolution: iterative  schemes}  
\blk

It has been seen that a simple replacement of a generator by a nonlinear function thereof cannot lead to an improved scaling of the average estimation error, in marked contrast to the  situation regarding the  local precision.  However, a careful analysis shows that  with  a suitable sum of nonlinear generators,  the bounds in Eqs.~(\ref{ebound}) and (\ref{inf}) allow for an  enhanced scaling of $\epsilon(\hat{\Phi})$, and that this enhanced scaling  could,  plausibly,  be achievable by adaptive measurements.  
This resolves the above \blu puzzle \blk to some degree.  Significantly, however,   the  scaling of the average estimation error does not necessarily achieve the same scaling as the local precision. 

As a first example, let  $G({ l})$ denote the nonlinear generator $(J_z)^2$, for ${ l}$ qubits, and let $\rho({ l})$ denote an equally weighted superposition of two eigenstates of $G({ l})$, corresponding to its minimum and maximum eigenvalues (i.e., to 0 and ${ l}^2$ if ${ l}$ is even, and $1$ and ${ l}^2$ if ${ l}$ is odd). Now consider the total generator and corresponding composite probe state defined by
\begin{equation} \label{adapt}
 G_{\rm  it}  := \sum_{{ k}=1}^{ K} G(n_{ k})\,,~~~~\rho_0:=\otimes_{{ k}=1}^{ K} \,\rho(n_{ k})\,,
\end{equation}
with $n_{ k}:=\lceil 2^{({ k}-1)/2}\rceil$ (where $\lceil x\rceil$ denotes the smallest integer not less than $x$).
Since the phase shift generated by $G({ l})$ has period $\approx 2\pi/{ l}^2$, this ensures that the phase shift generated by $G(n_{ k})$, on the probe state component $\rho(n_{ k})$, has period $\approx (2\pi)/2^{{ k}-1}$.  

The  basic idea is that the ${ k}$th bit in a binary expansion of  $\Phi/(2\pi)$ is estimated from the ${ k}$th component of the probe state.  Note that it is impossible to obtain more than 1 bit from each component of the probe state, \blu i.e., more than $\ln 2$ nats, \blk as a consequence of Eq.~(\ref{inf}) and the property $A_{G({ l})}(\rho(k))=H(G({ l})|\rho({ l}))=\ln 2$.  This is the idea behind the famous quantum phase estimation algorithm \cite{nielsen} for linear phase shifts, 
subsequently generalized in \cite{higgins,berry,Higgins09}. In practice, to achieve 
the best scaling for the average estimation error it  may be  necessary to use ${ M} > 1$ copies of each component of the probe state to estimate each bit accurately, when combined with an adaptive measurement sequence \cite{higgins,berry}. Counter-intuitively, it is the least significant (${ K}$th) bit of $\Phi/(2\pi)$ that should be determined first, to allow the optimal measurement of the $({ K}-1)$th bit, and so  on up  to the most significant bit. 

The   total number of qubits required in the above setup is $n={ M}\sum_{ k} n_{ k}  \approx   M  (2^{{ K}/2}-1)/(\sqrt{2}-1)$. 
Further, there are $2^{ K}$ distinct eigenvalues of $G$, of the form $\approx \sum_{ k} b_{ k}2^{{ k}-1}$ for $b_{ k}=0$ or 1, where these have a uniform distribution for $\rho_0$. 
Taking into account the ${ M}$ copies of $G_{\rm it}$  and $\rho_0$,  the corresponding generator $G=G_{\rm it}^{(1)}+\dots+G_{\rm it}^{({ M})}$ has eigenvalues ranging from $0$ to ${ M}(2^{ K}-1)$, implying a $G$-asymmetry $A_G(\otimes^{ M}\rho_0)  \leq  H(G|\otimes^{ M}\rho_0)\leq \ln [{ M}2^{ K}]$ (note  that for $M>1$ the distribution of $G$ over $\otimes^r \rho_0$ is not uniform, and this upper bound is not tight).  
 Thus  Eq.~(\ref{ebound}) yields the following 
lower bound for the average estimation error,  
\begin{equation} \label{adquad}
\epsilon(\hat{\Phi})\geq \frac{(2\pi/e)^{1/2}}{{ M}[1+(\sqrt{2}-1)n/{ M}]^2} \approx \frac{(2\pi/e)^{1/2}}{(3-2\sqrt{2})} \frac{{ M}}{n^2} ,
\end{equation}
in the worst-case scenario of a completely random phase shift.
Note 
that this bound is compatible with the scaling expected for a scheme that determines the first $K$ bits of $\hat\Phi$, 
giving $\epsilon(\hat{\Phi})\leq (2\pi)/2^{ K+1} \sim  (M/n)^2$. In other words, for $M$  large enough  for this 
bitwise estimation scheme to work, we would expect the scaling with $n$ in \erf{adquad} to be attainable. 

Thus, this adaptive scheme demonstrates the possibility of an asymptotic $n^{-2}$ scaling for the average estimation error.  This is the same scaling (up to a constant factor) as for the optimal local precision for the generator $G(n)$ \cite{boixo1, boixo3}. Furthermore, an $n^{-{ q}}$ asymptotic scaling can be obtained for an analogous adaptive scheme based on the nonlinear generator $(J_z)^{ q}$, with  
\begin{equation} \label{adk}
\epsilon(\hat{\Phi}) \gtrsim \frac{c_{ q} { M}^{{ q}-1}}{n^{ q}} \,
\end{equation} 
for a phase shift random over $[0,2\pi)$, where $c_{ q}$ increases exponentially with ${ q}$. Analogous results may be obtained for  iterative  implementations of the schemes  in \cite{luis1,luis2}.

However, a correspondence of scalings between $P_\phi(\hat{\Phi})$ and $\epsilon(\hat{\Phi})$ does not hold more generally.  For example, let  $G({ l})$ instead denote the nonlinear generator $H+2^{{ l}-1}$ for ${ l}$ qubits, with $H$ as in Eq.~(\ref{ha})  but with ${ l}$ in place of $n$, and with the additive constant being  chosen to simplify eigenvalue counting.  Further, let $\rho({ l})$ denote an equally weighted superposition of the two eigenstates corresponding to the minimum and maximum eigenvalues $0$ and $2^{{ l}}$ of $G({ l})$  (note such superpositions include the separable states $\Ket{z,z,\dots,z}$ and $\Ket{-z,-z,\dots,-z}$ \cite{roy,royeig}).   
Successive estimation of ${ l}$ binary digits then corresponds to  $n_{ k}={ k}-1$  in Eq.~(\ref{adapt}), requiring a total qubit number  $n={ M}{ K}({ K}-1)/2\approx { M}{ K}^2/2$.  One has   $A_G(\otimes^{ M}\rho_0)\leq  \ln ({ M}2^{ K})$ as before,  yielding the lower bound
\begin{equation} \label{adroy}
\epsilon(\hat{\Phi})\gtrsim (2\pi/ e)^{1/2}{ M}^{-1}\, 2^{-\sqrt{2n/{ M}}}
\end{equation}
for the case of a completely random phase shift.  
 Thus the scaling with $n$ is  considerably worse than  the $2^{-n}$ scaling of the local precision for the corresponding 
 single generator  scheme  \cite{roy}.  

This last result demonstrates that the local precision does not necessarily characterise the performance of the average estimation error even for adaptive implementations. It follows that comparisons between various schemes, whether linear or nonlinear, should be made on the basis of the operationally significant quantity, $\epsilon(\hat{\Phi})$ in Eq.~(\ref{error}), rather than $P_\phi(\hat{\Phi})$ in Eq.~(\ref{dis}).

\section{ Linear schemes with optimal scaling}


\subsection{Probes comprising $n$ qubits}

Since any generator $G$ for $n$ qubits has at most $2^n$ distinct eigenvalues,  it follows from Eqs.~(\ref{pure}) and (\ref{bounds}) that $A_G(\rho_0)\leq H(G|\rho_0)\leq n\ln 2$, with equality for a probe state that is an equally weighted superposition of the corresponding eigenstates.  Hence, from Eq.~(\ref{ebound}), the best possible scaling for the average estimation error satisfies
\begin{equation} \label{fun1}
\epsilon(\hat{\Phi})  \geq (2\pi e)^{-1/2}\, e^{H(\Phi)}\,2^{-n}  .
\end{equation}
Estimation schemes having an exponential scaling linear in $n$ are therefore of fundamental interest. Note, as per Eq.~(\ref{adroy}), that such a scaling is {\it not} attained via an adaptive implementation of the nonlinear scheme in \cite{roy}, despite  the local precision scaling as $2^{-n}$ for this scheme.

It is possible that nonlinear schemes exist with an exponential scaling in $n$ for $\epsilon(\hat{\Phi})$.   Here we show that, surprisingly, such a scaling can be achieved with {\em linear} generators. The extra ingredient that makes this possible is to allow for  multiple (and varied)  applications of the phase shift  
prior to measurement.  

In particular, following the ideas in Higgins {\it et al.} \cite{higgins}, consider a probe system comprising $m$ unentangled qubits, each in the state $|+\rangle=(\ket{-z}+|z\rangle)/\sqrt{2}$, where the ${ k}$th qubit is subjected to  $2^{{ k}-1}$  applications of a linear phase shift generated by $(1+\sigma^{({ k})}_z)/2$.  In Ref.~\cite{higgins} this was achieved experimentally via multiple passes through a medium.  Another possibility would be to suitably increase interaction times of the qubits with the phase shift medium.  
The total generator and the probe state therefore have the forms
\begin{equation} \label{multi}
G_{\rm  it} = \sum_{{ k}=1}^{ K} 2^{{ k}-1}  (1+\sigma^{(j)}_z) /2 ,~~~~\rho_0 = \otimes^{ K} |+\rangle\langle +| .
\end{equation}
The total phase shift of the ${ k}$th qubit thus has period $2\pi/2^{{ k}-1}$, and so can be used to estimate the ${ k}$th bit of $\Phi/(2\pi)$, in the  adaptive   manner explained above \cite{higgins}.
 
If, as for the nonlinear schemes above, ${ M}$ copies are used to estimate each bit accurately, then the total number of qubits required is $n={ M}{ K}$.   Following the style of arguments used in the preceding section gives a lower bound on the 
 the average estimation error  of 
 \begin{equation} \label{fun1.5}
\epsilon(\hat{\Phi})  \geq (2\pi e)^{-1/2}\, e^{H(\Phi)}\,{ M}^{-1}\,2^{-n/{ M}}  .
\end{equation}
 In this case it would appear that minimizing ${ M}$ could make a big improvement to the precision, and for ${ M}=1$  
 the ultimate scaling (\ref{fun1}) might be achievable. However it must be remembered that Eq.~(\ref{fun1.5}) 
 is merely a lower bound. Moreover, from the arguments in the preceding section, 
 it is only for  $M$ sufficiently large that we expect these bitwise estimation schemes to attain the scaling with $n$ 
 of the lower bounds.
 
 Luckily,   in this case, 
we  can compare this bound  to the actual performance of the best known adaptive schemes, 
 for an initially completely random phase,  
as this has been studied extensively.  
These studies were done using the Holevo variance $V_H(\hat{\Phi})$ \cite{WisKil97} rather than the 
 average estimation error, but when these are small (as here, for $n$ large), $V_H(\hat{\Phi})\leq \epsilon(\hat{\Phi})^2 \leq (\pi/2)^2 V_H(\hat{\Phi})$ \cite{BWB}. 
 In terms of scaling with $n$, the best performances is  indeed  for ${ M}=1$, which corresponds to the quantum phase estimation algorithm \cite{nielsen} and yields \cite{berry}
 \begin{equation} \label{r1}
 \epsilon(\hat{\Phi}) = c \times 2^{-n/2}.
 \end{equation}
(The constant $c \approx 1.18$ can be evaluated by performing the integral of the distribution of phase estimates, 
 Eq.~(4.5) of \cite{berry}.) Although this is not identical scaling to (\ref{fun1}), it is still exponential in $n$, unlike (\ref{adroy}).   
 For ${ M}=2$, $3$,  and $4$,  the performance  scales as $2^{-n/4}$, while for ${ M}\geq 4$ it scales as $2^{-n/{ M}}$,  
 achieving the lower bound scaling in Eq.~(\ref{fun1.5})  as expected for ${ M}$ sufficiently large. 


 Note that in terms of ${\cal  N} = { M}(2^{{ K}+1}-1)$, the number of qubit-passes through the phase shift (which is the resource considered in \cite{higgins,berry,Higgins09,rapcom}), the change in scaling as ${ M}$ increases appears quite different. For ${ M}=1$ and ${ M}=2$, the scaling is ${\cal  N}^{-1/2}$; for ${ M}=3$  it is ${\cal  N}^{-3/4}$; and for ${ M}\geq 4$ it is ${\cal  N}^{-1}$. We emphasize that counting resources as above, in terms of the number of qubits $n$,  is necessary to enable consistent comparison with the nonlinear schemes considered in \cite{boixo1,ramsey,roy,boixo2,boixo3}. 
 Recently, some other papers 
 have also considered the number of qubits (or, more strictly, the number of qubit measurements) as a resource \cite{serge,ferrie}.  
However, as these were motivated by qubit gate characterization in solid-state quantum computing,  they imposed the constraint that 
the qubit measurement basis is fixed. In this case
 the only thing that can be chosen adaptively is the number of times the phase shift is applied to a given qubit prior to measurement. 
 In \cite{serge} numerical evidence was presented that, using a locally optimal (``greedy'') adaptive algorithm, a scaling 
 of approximately $2^{-0.1 n}$ is achievable. In \cite{ferrie} an analytical argument was given suggesting that 
 a scaling  of approximately $2^{-0.16 n}$ should be achievable. Neither achieves the scaling (\ref{r1}) of  
 schemes that allows adaptive controlled qubit rotations prior to measurement.  
%
%

\subsection{Probes comprising optical modes} 

For a optical probe containing $m$ orthogonal modes, let $N_m$ denote the photon number of the $m$th mode,  and $N$ denote the total photon number $N_1+\dots+N_m$.  The entropy of any generator $G=f(N_1,\dots,N_m)$ is bounded above by the joint entropy of $N_1,\dots,N_m$ (since the distribution of $G$ is a coarse graining of the joint distribution), yielding  via Eq.~(\ref{bounds})
\begin{eqnarray} \nonumber
A_G(\rho)&\leq&  H(G|\rho)\leq H(N_1,\dots,N_m |\rho) \\ 
&\leq& m\ln \left[ 1+\frac{\langle N\rangle}{m} \right]  +
\langle N\rangle \ln \left[ 1+\frac{m}{\langle N\rangle}\right]\\ 
&\leq&  \ln e^{m+\langle N\rangle} 
\end{eqnarray}
The second line follows from standard statistical mechanics techniques, and the third line from  the monotonic convergence of $(1+x/y)^y$ to $e^x$ as $y$ increases. 

Using Eq.~(\ref{ebound}), the average estimation error therefore has the fundamental lower bound
\begin{equation} \label{fun2}
\epsilon(\hat{\Phi})  \geq (2\pi e)^{-1/2}\, e^{H(\Phi)}\, e^{-m}\,e^{-\langle N\rangle}  \, ,
\end{equation}
for any generator which is a function of $N_1,\dots,N_m$.  This includes the total photon number, $N=N_1+\dots+N_m$, in particular \cite{yuen92}, but also includes, for example, the nonlinear generators $N^2$ and $(N_1)^2+\dots (N_m)^2$.  Similarly, using Eq.~(\ref{inf}), one has the fundamental upper bound  $m+\langle N\rangle$ 
 for the mutual information $H(\hat{\Phi}:\Phi)$.

It follows that estimation schemes with an exponential scaling linear in the average photon number are of fundamental interest.  
Further, a linear scheme, analogous to the one above for $n$ qubits, is sufficient to obtain such a scaling  (though with a different coefficient).   
In particular, the  ${ M}=1$  linear multipass scheme of Higgins {\it et al.} \cite{higgins}, equivalent to the quantum phase estimation algorithm \cite{nielsen}, is precisely such a scheme,    involving $m =  2{ K}$  modes, with each pair of modes in the  superposition  state $(|0\rangle|1\rangle-|1\rangle|0\rangle)/\sqrt{2}$.  
Hence,  ${ K} = \langle N\rangle=m/2$, and from \erf{r1},  $\epsilon(\hat{\Phi}) \simeq 1.18 \times e^{ -  K  \ln(2) / 2 }$ asymptotically, 
which is consistent with  Eq.~(\ref{fun2}) as here  $m+\langle N\rangle = 3 { K}$.  
Again, it should be noted, as for the qubit case above, the measure of resources considered here is the total mean photon number $\langle N\rangle$ required for the scheme, rather than   the number of photon-passes ${\cal N}$ through the phase shift medium as in Ref.~\cite{higgins}. 
 Again, we use $\langle N\rangle$ to enable comparison with  various linear and nonlinear estimation schemes \cite{luis1,luis2}. Moreover,   photon number is a  natural measure of interest to consider,  as it characterises the energy resources required for a given optical  scheme.  

\section{Discussion}

The average estimation error and mutual information have been shown to satisfy the general entropic bounds in Eqs.~(\ref{inf}) and (\ref{ebound}), for any shift generator having a discrete spectrum, \blu and for any prior distribution of the shift parameter. \blk  While, for phase shift generators,  the $G$-asymmetry  can be bounded  above  in terms of the variance of the generator,  via Eq.~(\ref{strict}), the $G$-asymmetry is typically much less than this upper bound.  Hence,  the average estimation error can scale very differently to the local precision  in Eq.~(\ref{dbound}), in terms of available resources such as number of qubits or total input photon number.  

Indeed, somewhat \blu surprisingly, \blk a simple replacement of a linear generator by some nonlinear function thereof may have no effect on the average estimation error, yet lead to a marked improvement of the local precision.  Further, while such scaling differences can disappear for iterative estimation schemes, this is not always the case. 

 \blu It follows that the optimal scaling of the local precision, for a given value of the shift parameter, should be treated with some caution.  As noted in relation to Eq.~(\ref{unbiased}), the local precision is a direct measure of the average root mean square error only over an interval for which the corresponding estimate is (approximately) unbiased.  However, \grn many estimators in the literature are unbiased only over a very small interval, similar \blu in magnitude to the local precision itself.  In such a case \mg (which is relevant, for example, in phase tracking and phase sensing applications \cite{durk,xiang}), \blk for the optimal scaling to be achievable, the amount of prior information required about the shift parameter is typically so great that the estimate itself can only extract up to 1 bit of further information, irrespective of the number of resources $n$.  \grn Examples of this phenomenon \blu have been given in Secs.~III~C and IV.
 
It is concluded \blu from the above \blk that \gold meaningful \blk comparisons between various estimation schemes \gold are most easily  \blk made on the basis of the operationally significant quantity, $\epsilon(\hat{\Phi})$ in Eq.~(\ref{error}), rather than $P_\phi(\hat{\Phi})$ in Eq.~(\ref{dis}). \grn Alternatively,  \gold if $P_{\phi'}(\hat{\Phi})$ \mg for some $\phi'$ \gold is used, then it \blk should be supplemented by the interval over which this estimate is (approximately) locally unbiased, \mg i.e., the interval for which the local precision of the estimate corresponds to the actual root-mean-square error of the estimate, $\langle (\hat{\Phi}-\phi)^2\rangle_{\phi}$. \blk \mg Note that the width of this interval also bounds the width of any `sweet spot' for which $P_\phi(\hat{\Phi})=P_{\phi'}(\hat{\Phi})$, and hence bounds the width of the prior phase probability density required to ensure a precision of $P_{\phi'}(\hat{\Phi})$ is actually achieved via measurement.  As per Eq.~(\ref{unbiased}), the \gold estimation error averaged over this prior distribution \mg will then (approximately) be equal to  $P_{\phi'}(\hat{\Phi})$. 

Universal lower bounds for the average estimation error have also been given, in terms of the number of qubits or photons available, in Eqs.~(\ref{fun1}) and (\ref{fun2}). \blu Like Eq.~(\ref{ebound}), \blk these bounds are independent of the form of the generator, and hence apply equally well to both linear and nonlinear schemes,  including multipass schemes.  They imply that it is impossible to achieve a scaling better than exponential, in terms of qubit number, and in terms of  input photon number  plus number of modes, respectively.   Further,  exponential scalings  {\it can}  be attained by multipass  linear estimation schemes, as has indeed  been  shown experimentally  for optical phase shifts \cite{higgins}. 
It follows that, in terms of the best possible scaling that can be achieved relative to qubit or input photon number, nonlinear schemes offer no fundamental advantage over linear schemes \grn if multiple passes are possible. \blk 

However, a practical advantage of nonlinear schemes may be a greater robustness to loss.   For example, multipass linear schemes of the type discussed above will be highly sensitive to loss due to multiple (or longer) interactions with the phase shift medium.  Thus it would be interesting  to find alternative physical implementations of the generator in Eq.~(\ref{multi}) and its optical analogue.  Further, \mg while only shot-noise scaling is achievable \blk for simple linear schemes in the presence of loss \cite{glm2004,knysh,escher}   (including  for the average estimation error \cite{nair}), there is evidence this may not be the case for nonlinear schemes \cite{luis2}.  Hence further investigation is required, including the determination of fundamental scaling bounds for lossy schemes analogous to Eqs.~(\ref{fun1}) and (\ref{fun2}).

 It would also be of interest to investigate the degree to which results generalise to the case of a  generator with a continuous spectrum,  such as  spatial  translations generated by a momentum operator. For example, the fundamental bounds (\ref{fun1}) and (\ref{fun2}) are universal, since $n\ln 2$ and $(m+\langle N\rangle)\ln e$ are respective upper bounds for mutual information, following via the Holevo bound. The possibility of other generalisations is supported by results such as a universal Heisenberg-type scaling for the average estimation error, in terms of $\langle |G|\rangle$, which holds for both discrete {\it and} continuous generators \cite{hallwise}.
 Further,  weaker measurement-dependent  entropic  bounds on the average estimation error, given in \cite{hallwise}, may prove helpful. Note that some differences are to be expected regarding linearity vs nonlinearity for the continuous case, since, for example, the property $H(f(G)|\rho)\leq H(G|\rho)$ no longer holds in general.




\acknowledgments
We thank D.W. Berry, R. Nair, J.A. Vaccaro and M. Zwierz for helpful discussions. This work was supported by the ARC Discovery Project DP0986503.


\begin{thebibliography}{0}


\bibitem{glm2004} 
 {V. Giovanetti, S. Lloyd and L. Maccone}, {\it Advances in quantum metrology},
  {Nature Photonics} {\bf 5}, {222} (2011).
\bibitem{helstrom} 
 {C. W. Helstrom}
{\it Quantum Detection and Estimation Theory}
 {(Academic Press, New York, 1976)}.

\bibitem{holevo}  
 A. S.  Holevo, {\it Probabilistic and Statistical Aspects of Quantum Theory} (North-Holland, Amsterdam, 1982). 

\bibitem{WisMil10} 
 {H. M. Wiseman  and G. J. Milburn},
{\it Quantum Measurement and Control}
 {(Cambridge University Press, Cambridge, 2010)}.

 \bibitem{luis1}
  {A. Luis}, {\it Nonlinear transformations and the Heisenberg limit},
  {Phys. Lett. A} { \bf 329}, {}{8} (2004).
 \bibitem{luis2} 
  {J. Beltran and A. Luis}, {\it Breaking the Heisenberg limit with inefficient detectors},
  {Phys Rev. A} { \bf 72}, {}{045801} (2005).
 \bibitem{boixo1} 
  {S. Boixo, S. T. Flammia, C. M. Caves  and J. M. Geremia}, {\it Generalized Limits for Single-Parameter Quantum Estimation},
  {Phys. Rev. Lett.} { \bf 98}, {}{090401} (2007).
 \bibitem{ramsey} 
   {S. Choi and B. Sundaram}, {\it Bose-Einstein condensate as a nonlinear Ramsey interferometer operating
beyond the Heisenberg limit},
  {Phys. Rev. A}{ \bf 77} {}{053613} (2008).
  \bibitem{roy} 
    {S. M. Roy and S. L. Braunstein}, {\it Exponentially Enhanced Quantum Metrology},
  {Phys. Rev. Lett.} { \bf 100}, {}{220501} (2008).
 \bibitem{boixo2}  
   S. Boixo,  A. Datta, M. J. Davis, S. T. Flammia, A. Shaji and C. M. Caves, {\it Quantum Metrology: Dynamics versus Entanglement},
  {Phys. Rev. Lett.} { \bf 101}, {}{040403} (2008).
 \bibitem{boixo3} 
  S. Boixo, A. Datta, S. T. Flammia, A. Shaji, E. Bagan and C. M. Caves, {\it Quantum-limited metrology with product states},
  {Phys. Rev. A} { \bf 77}, {}{012317} (2008).
 
   \blu
  \bibitem{nap1} M. Napolitano and M. W. Mitchell,  {\it Nonlinear metrology with a quantum interface}, New J. Phys. {\bf 12} 093016 (2010).
  
  \bibitem{nap2} M. Napolitano, M. Koschorreck, B. Dubost, N. Behbood, R. J. Sewell and M. W. Mitchell,  {\it Interaction-based quantum metrology showing
scaling beyond the Heisenberg limit}, 
  Nature {\bf 471}, 486 (2011).
\blk
 
  \bibitem{higgins} 
   B. L. Higgins, D. W. Berry, S. D. Bartlett, H. M. Wiseman and G. J. Pryde, {\it Entanglement-free Heisenberg-limited phase
estimation}, 
  {Nature} {\bf 450}, {}{393} (2007).
  
  \bibitem{berry} 
   D. W. Berry,  B. L. Higgins, S. D. Bartlett, M. W. Mitchell, G. J. Pryde and H. M. Wiseman, {\it How to perform the most accurate possible phase measurements}, 
  {Phys. Rev. A} {\bf 80}, {}{052114} (2009).
  

  \bibitem{Higgins09} 
   B. L. Higgins, D. W. Berry, S. D. Bartlett, M. W. Mitchell, H. M. Wiseman and G. J. Pryde, {\it Demonstrating Heisenberg-limited unambiguous
phase estimation without adaptive measurements}, 
  {New J. Phys.} {\bf 11}, {}{073023} (2009).

   \bibitem{rapcom} 
   {M. J. W. Hall,  D. W. Berry, M. Zwierz and H. M. Wiseman},    {\it Universality of the Heisenberg limit for estimates of random phase shifts}, 
  {Phys. Rev. A} {\bf 85}, {}{041802(R)} (2012).
  
  \bibitem{tsang}  {M. Tsang},  {\it Ziv-Zakai Error Bounds for Quantum Parameter Estimation}, Phys. Rev. Lett. {\bf 108}, 230401  (2012). 

 \bibitem{vantrees} 
  {H. L. Van Trees},
{\it Detection, Estimation and Modulation Theory Part 1}
{(Wiley, New York, 2001)}, Sec.~{2.4}.


\bibitem{refnote}   For the case of phase estimates, the range of the  error $\hat{\Phi}-\Phi$ is defined to be $[-\pi,\pi)$ to remove the phase reference ambiguity on the unit circle,  implying that  $0\leq \epsilon(\hat{\Phi})\leq\pi$.


 \bibitem{braun} 
   {S. L. Braunstein  and C. M. Caves}, {\it Statistical Distance and the Geometry of Quantum States}, 
  {Phys. Rev. Lett.} { \bf 72}, {}{3439} (1994).
 \bibitem{annphys}  
  S. L.  Braunstein, C. M.  Caves and G. J.  Milburn,  {\it Generalized Uncertainty Relations: Theory, Examples, and Lorentz Invariance}, Ann.  Phys.  (NY) {\bf 247}, 135 (1996). 
 
 \bibitem{unbiased} An estimate is locally unbiased, for a fixed phase shift value $\Phi=\phi$, if $\langle \hat{\Phi}\rangle = \phi$ and $d\langle \hat{\Phi}\rangle/d\phi=1$.
 
 \bibitem{future}  
  {D. W. Berry, M. J. W. Hall,  M. Zwierz and H. M. Wiseman}, {\it Optimal Heisenberg-style bounds for the average performance of arbitrary phase
estimates}, in preparation.

\mg 

\bibitem{durk} G. A. Durkin and J. P. Dowling,  {\it Local and Global Distinguishability in Quantum Interferometry}, Phys. Rev. Lett. {\bf 99}, 070801 (2007).


\bibitem{xiang} G. Y. Xiang,  B. L. Higgins, D. W. Berry, H. M. Wiseman and G. J. Pryde, {\it Entanglement-enhanced measurement of a completely unknown optical phase},  Nature Photonics {\bf 5}, 43 (2011).
 
 \blk
 
 \bibitem{hallwise} 
   {M. J. W. Hall  and H. M. Wiseman}, {\it Heisenberg-style bounds for arbitrary estimates of
shift parameters including prior information}, 
  {New. J. Phys.} { \bf 14}, {}{033040} (2012).
  
\bibitem{nair}  
  {R. Nair}, {\it Fundamental limits on the accuracy of optical phase estimation from rate-distortion theory}, 
 Eprint arXiv:1204.3761v1 [quant-ph].
 
 
 
 \bibitem{vacc1} J. A. Vaccaro, F. Anselmi, H. M. Wiseman, and K. Jacobs,  {\it Tradeoff between extractable mechanical work, accessible entanglement, and ability to act as a reference system, under arbitrary superselection rules}, Phys. Rev. A {\bf 77}, 032114 (2008).
 
 \bibitem{nielsen}  
   {M. A. Nielsen and I. L. Chuang},
{\it Quantum Computation and Quantum Information}
{Cambridge University Press, Cambridge, 2000}.

 \bibitem{lloyd}  S. Lloyd, Phys. Rev. A  {\it Quantum-mechanical Maxwell's demon}, {\bf 56},  3374 (1997).
 
 \bibitem{vacc2} J. A. Vaccaro,  {\it Group Theoretic Formulation of Complementarity}, in {\it Proceedings of the 8th International Conference on Quantum Communication, Measurement and Computing},  edited by O. Hirota, J.H. Shapiro, and M. Sasaki (NICT, Tokyo, 2006), pp.~421-424.
 \bibitem{spekk}  G. Gour, I. Marvian and R. W. Spekkens,  {\it Measuring the quality of a quantum reference frame: The relative entropy of frameness}, Phys. Rev. A {\bf 80}, 012307 (2009).
 \bibitem{vacc3} J. A. Vaccaro, {\it Particle-wave duality: a dichotomy between
symmetry and asymmetry}, {Proc. Roy. Soc. A} {\bf 468}, 1065 (2012).
 
\bibitem{yuen92}
  {Yuen H. P.},  {\it The Ultimate Quantum Limits on the Accuracy of Measurements}, in
 {\it Proceedings of the Workshop on Squeezed States and Uncertainty
Relations},
NASA Conference Publication 3135
 {(Goddard Space Flight Center, MD, 1992)}
pp.~{13}-{21}
(also at 
http:$//$ntrs.nasa.gov$/$archive $/$nasa$/$casi.ntrs.nasa.gov$/$19920012804$\underline{~}$
 1992012804.pdf).
 
 \blu
 \bibitem{cover} T. M. Cover and J. A. Thomas, {\it Elements of Information Theory, 2nd edn} (Wiley, New Jersey, 2006), Corollary to Thm~8.6.6.
 \blk
 
 \bibitem{rate} 
  {P. Harremo\"{e}s},
{\it Probability and information -  Occam's  razor in action},
in preparation (see also http:$//$ www.math.ku.dk$/$$\sim$moes$/$bog4.pdf, Chap. 12).
 

\bibitem{disc}
 W. H. Mow,  {\it A tight upper bound on discrete entropy}, IEEE Trans. Inf. Theory {\bf 44}, 775 (1998).  

 \bibitem{noon}  P. Kok, H. Lee, and J. P. Dowling,  {\it Creation of large-photon-number path entanglement conditioned on photodetection}, Phys. Rev. A {\bf 65}, 052104 (2002); D. Leibfried, M. D. Barrett, T. Schaetz, J. Britton, J. Chiaverini, W. M. Itano, J. D. Jost, C. Langer and D. J. Wineland
,  {\it Toward Heisenberg-Limited Spectroscopy with Multiparticle Entangled States}, Science {\bf 304}, 1476 (2004).

\bibitem{royeig} It is straightforward to verify that  $(\Ket{z,\dots,z} \pm \Ket{-z, \dots,-z})/\sqrt{2}$ and $(\Ket{z,\dots,z} \pm i \Ket{-z, \dots,-z})/\sqrt{2}$  are eigenstates of $H$ and $A$ respectively, in Eq.~(\ref{ha}), with eigenvalues $\pm 2^{n-1}$.  Hence, since the sum of the squares of the eigenvalues of $H$ and $A$ is given by ${\rm tr}[H^2+A^2] = {\rm tr}[(H+iA)(H+iA)^\dagger] = \{ {\rm tr}[(\sigma_x+ i\sigma_y)(\sigma_x -i\sigma_y)]\}^n = 4^{n}$, these eigenvalues are nondegenerate and all remaining eigenvalues vanish. 

 \bibitem{WisKil97}
{H. M. Wiseman  and R. B. Killip}, {\it Adaptive single-shot phase measurements: A semiclassical approach}, 
  {Phys. Rev. A} {\bf 56}, {944} (1997).
  
\bibitem{BWB}  D. W. Berry, H. M. Wiseman and J. K. Breslin,  {\it Optimal input states and feedback for interferometric phase estimation}, Phys. Rev. A {\bf 63}, 053804  (2001).
  
 \bibitem{serge} 
   {A. Sergeevich, A. Chandran, J. Combes, S. D. Bartlett, and H. M. Wiseman}, {\it Characterization of a qubit Hamiltonian using adaptive measurements in a fixed basis}, 
  {Phys. Rev. A} {\bf 84}, {}{052315} (2011).
 
 \bibitem{ferrie}
 {C. Ferrie, C. E. Granade  and D. G. Cory}, {\it How to best sample a periodic probability distribution, or on the accuracy of Hamiltonian finding strategies}, 
 Eprint arXiv:1110.3067v1 [quant-ph].
 
 \mg
 \bibitem{knysh} S. Knysh, V. N. Smelyanskiy, and G. A. Durkin, {\it Scaling laws for precision in quantum interferometry and the bifurcation
landscape of the optimal state}, Phys. Rev. A {\bf 83}, 021804(R) (2011).
 \blk 
 
 \bibitem{escher} 
 {B. M. Escher, R. L. de Matos Filho  and L. Davidovich}, {\it General framework for estimating the ultimate precision limit in noisy quantum-enhanced metrology}, 
  {Nature Phys.} {\bf 7}, {}{406} (2011).
 
  
 
 

 
 


\end{thebibliography}
\end{document}